# From power to spectrum: an off-the-shelf photodiode turned into a spectrometer by its bias alone

Roi Yozevitch, Shmuel Sternklar, and Ziv Glasser

Department of Electrical and Electronic Engineering, Ariel University, Ariel 40700, Israel

## Abstract

Computational spectrometers replace dispersive optics with detectors whose spectral response can be varied, but every reported single-detector device obtains this response through dedicated fabrication: two-dimensional heterostructures, graded epitaxial junctions, organic stacks, filters or photonic circuits. We demonstrate a generic, fabrication-free alternative in which a standard commercial photodiode becomes a spectrometer through its electrical drive alone, requiring no modification of the device, no optical filter and no modulation of the light. The mechanism is bias-modulated optoelectronic chromatic dispersion (OED): light of different wavelengths is absorbed at different depths, and a sinusoidal bias changes how carriers generated at each depth are collected, so that every spectrum is encoded in the shape of the photocurrent waveform over one bias period. Applied to an unmodified InGaAs PIN photodiode at room temperature, a 10 μs waveform decoded by a neural network reconstructs spectra of one to three laser lines across the 80 nm of the telecom C+L band, locating the lines to 0.3–0.4 nm, counting them in about 99% of measurements and identifying their power levels, with line shifts tracked to tens of picometres. Because OED relies on universal semiconductor transport physics, the approach is material-agnostic and applies in principle to any off-the-shelf photodiode, turning detectors already deployed as power monitors into low-cost spectral sensors.

## 1. Introduction

Optical spectrometers are central to science, telecommunications and industry, but conventional instruments rely on gratings or interferometers and remain bulky and expensive. Computational spectrometers replace these elements with detectors whose spectral responses differ and recover the spectrum numerically from the measured signals [1–3]. One family places an array of spectral encoders in front of the detector—colloidal quantum dots [4], compositionally graded nanowires [5], plasmonic filters on InGaAs [6] or integrated photonic circuits read by an InGaAs photodiode [7,8]. A second family uses a single detector whose responsivity is tuned electrically and samples the photocurrent at a series of bias or gate voltages [9,10]. The tuning has been obtained by shifting the absorption edge of black phosphorus through the Stark effect [9], by tuning the band alignment of van der Waals heterojunctions and tunnel diodes [11–13], and by changing with bias which depth of the absorber contributes to the photocurrent, using a compositionally graded AlGaAs junction [14], an organic photodetector with an optical spacer [15] or tandem organic subcells [16]; the temporal photoresponse has served as an additional spectral signature [17,18].

These devices reach sub-nanometre peak-wavelength accuracy [10,14,16], but they share one requirement: the spectral selectivity is built into the detector by dedicated fabrication—exfoliated or stacked two-dimensional materials, custom-grown graded III–V junctions, organic multilayers, filters or photonic circuits. Such detectors are not available off the shelf, their spectral response is fixed by the fabrication process, and several have been characterized only at cryogenic temperature [9,11]. The ubiquitous component of optical systems—the commercial photodiode, produced in very large volumes and present in every optical receiver, power monitor and transceiver—has so far been used only to measure optical power, or as the read-out element behind an engineered encoder [7].

In this work we introduce a fundamentally different approach: instead of engineering a detector, we program an existing one. The approach exploits optoelectronic chromatic dispersion (OED)—the dependence of a photodiode's temporal response on wavelength, which arises because light of different wavelengths is absorbed at different depths and the photogenerated carriers therefore travel different distances and experience different fields before they are collected [19]. Our group has shown that OED appears as a wavelength-dependent RF phase and amplitude of the photocurrent under intensity-modulated light in germanium and silicon photodiodes and photovoltaic cells [19–21], and has used it for wavelength monitoring with femtometre resolution [22], for strain sensing in unmodified solar cells [23], and, combined with machine learning, for reconstructing one and two wavelengths from an intensity-modulated photodiode [24]. These schemes require the light under test to be modulated. Modulating the bias instead, bias-modulated OED encodes the wavelength while the optical input is left untouched [25].

Here we show that bias-modulated OED turns an unmodified commercial InGaAs PIN photodiode into a spectrometer. The device contains no structure engineered for spectral selectivity, the light is neither filtered nor modulated, and the only added elements are a function generator and a digitizer: a 100 kHz sinusoidal bias and one 10 μs photocurrent waveform. From this waveform a neural network reconstructs spectra of one to three laser lines across the telecom C+L band (1530–1610 nm), retrieving the wavelength, the number and the power level of each line. Because the spectral encoding relies only on absorption depth and carrier collection, properties shared by all semiconductor photodiodes, the approach is material-agnostic and fabrication-free. In the following section we describe the working principle; we then characterize the photoresponse, present the reconstruction of multi-line spectra, and discuss the performance, operation on later days, and the generality of the method.

## 2. Theory and working principle

**Optoelectronic chromatic dispersion.** Light entering a photodiode is absorbed with a wavelength-dependent coefficient $\alpha(\lambda)$, so the carriers are generated with a depth profile $G(z,\lambda)\alpha(\lambda)e^{-\alpha(\lambda)z}$. Carriers generated at different depths drift and diffuse over different distances, through regions of different electric field, before they reach the contacts. The photocurrent therefore responds with a delay $\tau(\lambda)$ and a collection efficiency $\eta(\lambda)$ that both depend on wavelength—the optoelectronic chromatic dispersion [19]. The effect is strongest just below the cutoff wavelength $\lambda_c$ of the absorber, where a small change of wavelength produces a large fractional change of the absorption length while the responsivity is still high [19,23]. For lattice-matched InGaAs ($\lambda_c$≈1.7 μm) the telecom C+L band lies in exactly this region: across 1530–1610 nm the absorption length grows continuously with wavelength, so that longer-wavelength light is absorbed deeper in the absorber (Fig. 1a).

**Bias modulation.** In earlier OED schemes the light is intensity-modulated and the dispersion is read as an RF phase shift. Here the light is left untouched and the bias is modulated instead. The bias sets the electric field and the extent of the depleted region, and therefore how efficiently and how quickly carriers generated at a given depth are collected. A sinusoidal bias $V(t)=V_0+V_1\sin\omega t$ ($V_0$=6.5 V, $V_1$=3.5 V, $\omega/2\pi$=100 kHz) sweeps the device through a continuum of collection conditions within each period, and each wavelength, with its own generation profile, follows a different photocurrent trajectory (Fig. 1b). The chain of effects is

$$\lambda \rightarrow \alpha(\lambda) \rightarrow G(z,\lambda) \rightarrow \eta(V(t),\lambda),\tau(V(t),\lambda) \rightarrow \Delta I(t;\lambda), \tag{1}$$

so that the spectrum is encoded in the shape of the photocurrent waveform. Depth-selective collection is also the principle of engineered single-pixel spectrometers, in which the wavelength-to-depth mapping is set by a graded bandgap [14] or an optical spacer [15]; here it is supplied by the intrinsic absorption of a standard absorber.

**Photoresponse vector.** For an incident spectrum $P(\lambda)$ the light-induced current at time $t$ within the bias period is

$$\Delta I(t) = I(t) - I_{\mathrm{dark}}(t) = \int R(V(t),\lambda;P)\, P(\lambda)\, d\lambda, \tag{2}$$

where $R$ is the responsivity of the photodiode at the instantaneous bias $V(t)$. The argument $P$ indicates that $R$ may itself depend on the incident optical power, which we find to be the case (Sec. 4.2). Because the bias is swept continuously rather than held at fixed values, $R$ describes the response during the sweep and need not equal the static responsivity at the same voltage. Sampling $\Delta I(t)$ at 8000 instants per period yields a photoresponse vector that plays the role of the responsivity matrix rows of static-bias spectrometers [9]. Since the bias is periodic, the waveform can equivalently be written as a sum of bias harmonics, $\Delta I(t)=\Sigma_k|F_k|\cos(k\omega t+\varphi_k)$. Operation follows three steps (Fig. 1c): *learning*, in which the response to known laser lines is recorded; *sampling*, in which one bias period of an unknown input is recorded; and *reconstruction*.

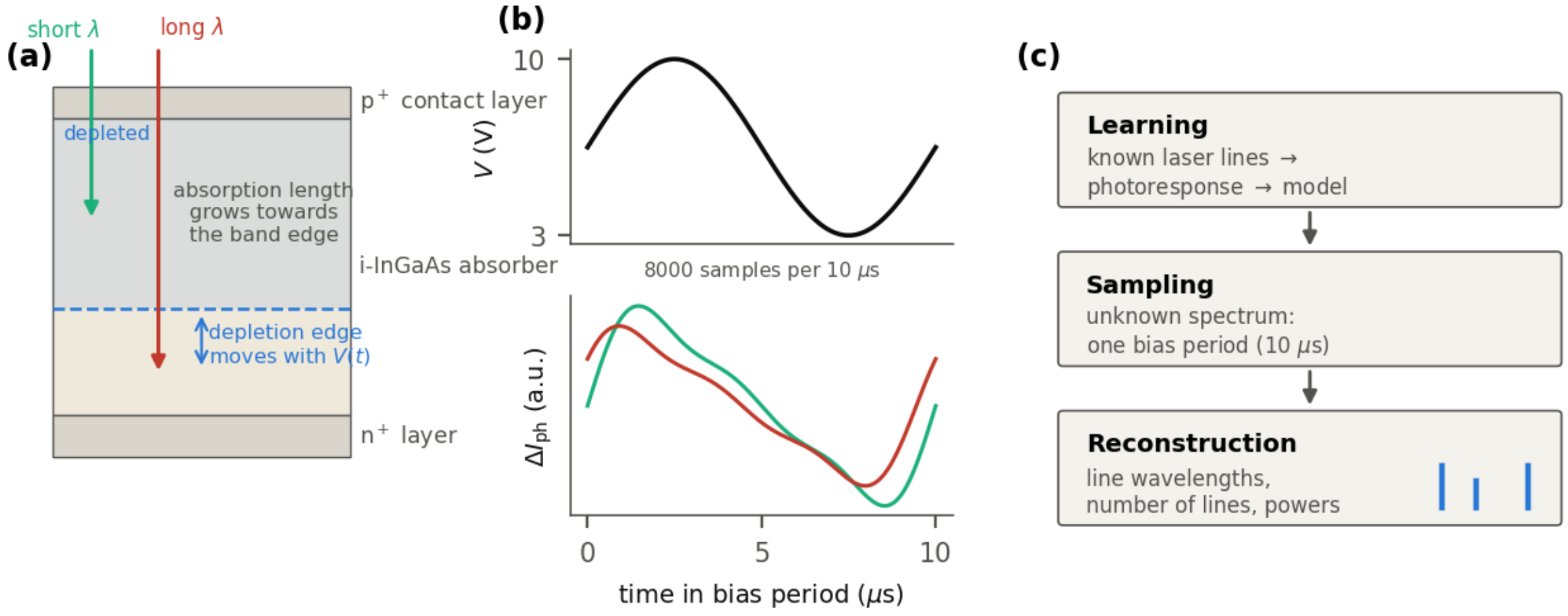

**Fig. 1.** Working principle of the bias-modulated OED spectrometer. (a) Light of longer wavelength is absorbed deeper in the InGaAs absorber; the bias modulates the collection of carriers generated at each depth (schematic, not to scale). (b) The sinusoidal bias $V(t)$ and, schematically, the light-induced photocurrent for two wavelengths: each wavelength produces its own waveform, sampled at 8000 points per 10 μs period. (c) Learning with known laser lines, sampling of an unknown spectrum within one bias period, and reconstruction of the line wavelengths, number and powers.

## 3. Experimental setup

Figure 2 shows the configuration. The detector is a standard commercial InGaAs PIN photodiode (JDSU, 2.5 GHz bandwidth), used as purchased. Light from up to three tunable lasers (1530–1610 nm) is combined in fiber; a fraction is monitored on an optical spectrum analyzer. A function generator applies the sinusoidal bias between 3 and 10 V. The photodiode is terminated in 100 Ω; a bias tee separates the DC photocurrent, recorded directly, from the AC component, which is amplified by a low-noise amplifier (1 kHz–1.8 GHz) and digitized by an oscilloscope as 8000 samples over one bias period, triggered by the function generator. Each measurement session begins with a dark reference (all lasers off), which is subtracted from every waveform of the session. Each laser is set to 4 or 8 mW, and simultaneous lines are kept at least 0.5 nm apart.

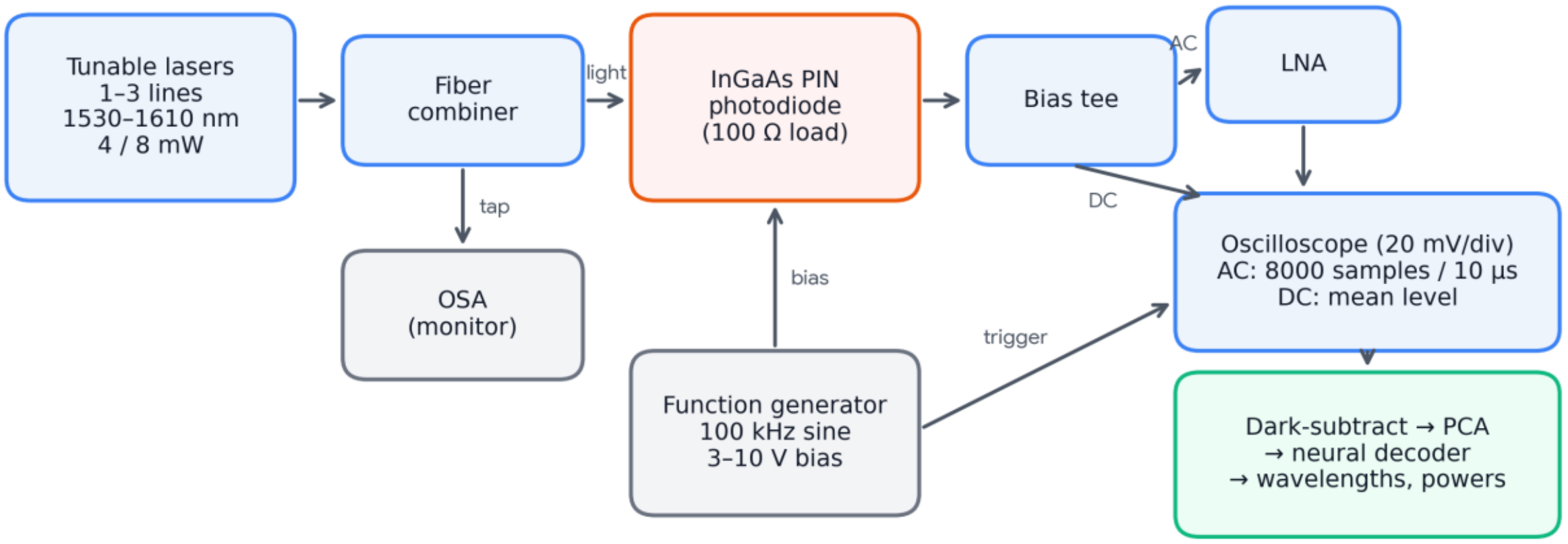

**Fig. 2.** Experimental configuration. Up to three tunable lasers illuminate the unmodified photodiode, which is driven by a sinusoidal bias; the output is split into DC and AC parts, and the AC waveform over one bias period is digitized and dark-subtracted.

The data set (Table 1) comprises 29 529 waveforms recorded in 20 sessions between March and July 2025: single lines on a 1-nm grid over 1530–1610 nm, two-line spectra (C-band pairs and C+L pairs), three-line spectra, and fine scans with 0.01 and 0.001 nm steps.

**Table 1.** Measured spectra (dark-subtracted waveforms with known set wavelengths and powers).

| Spectra | Waveforms | Sessions | Wavelengths and powers |
|---|---|---|---|
| Single line | 4 535 | 11 | 1530–1610 nm, 1-nm grid; 4 and 8 mW |
| Two lines | 14 750 | 12 | C-band pairs (one line on a 0.5-nm grid, $\Delta\lambda \geq 0.5$ nm) and C+L pairs (1-nm grid) |
| Three lines | 6 036 | 5 | $\lambda_1 \in \{1530, 1535, 1540\}$ nm, $\lambda_2$ 1571–1610 nm, $\lambda_3$ C-band |
| Fine single line | 2 204 | 2 | 1550–1551 nm in 0.01 and 0.001 nm steps |
| Fine two lines | 2 004 | 1 | $\lambda_1$=1550 nm, $\lambda_2$ 1550.5–1551.0 nm in 0.001 nm steps |
| Total | 29 529 | 20 | |

## 4. Results

### *4.1 The spectral fingerprint in the photocurrent waveform*

Figure 3(a) shows the light-induced AC photocurrent over one bias period for five wavelengths across the band. The waveform does not merely scale with wavelength: it changes shape, as expected when carriers generated at different depths respond differently to the bias sweep. The wavelength-dependent part of the response, mapped over wavelength and time in Fig. 3(b), varies smoothly and non-monotonically across the 80 nm band, so that the waveform as a whole forms a unique fingerprint for each wavelength. The fingerprint is concentrated where the bias changes fastest, near its mid-level crossings, and in the first eight bias harmonics (100–800 kHz); harmonic amplitudes carry more of it than phases. This is the signature of bias-modulated OED: the information appears where the collection conditions change most rapidly.

How many independent spectral channels does one bias period provide? Decomposing the wavelength response into principal components (Fig. 3c) shows that five components hold 99.96% of its variance. About 20 components carry wavelength information above the shot-to-shot noise, and the first five are also well above the variation between measurement days (signal-to-drift variance ratios of 602, 107, 33, 26 and 10). A single 10 μs waveform thus provides a compact set of robust spectral channels, generated by the device physics rather than by fabricated structures.

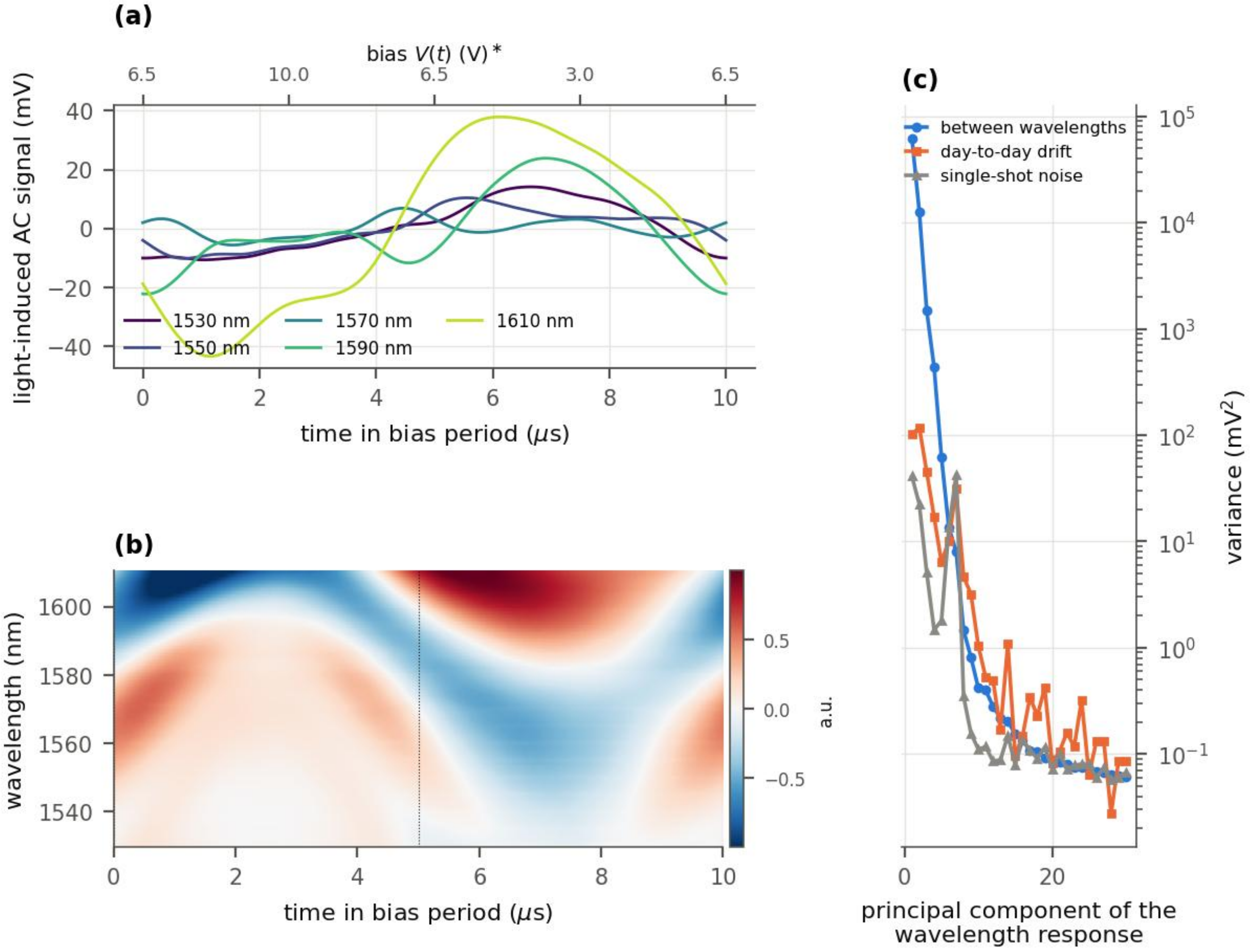

**Fig. 3.** The spectral fingerprint (single line, 4 mW, mean over 11 sessions; harmonics 1–12). (a) Light-induced AC signal over one bias period for five wavelengths; top axis: instantaneous bias, assuming the trigger coincides with the rising mid-level crossing of the bias. (b) Wavelength-dependent part of the response normalized by the DC photocurrent (deviation from the band mean) versus wavelength and time; dotted lines: bias mid-level crossings. (c) Variance along the principal components of the wavelength response: between wavelengths (signal), between sessions for the same wavelength (day-to-day drift) and between repeated shots (noise).

### *4.2 A nonlinear, line-coupled photoresponse*

Is the waveform of a multi-line spectrum simply the sum of the waveforms of its lines, as assumed by linear computational spectrometers? Each session records the single-line waveform of the first laser, and seconds later the two-line waveform obtained when the second laser is switched on. Their difference is the contribution of the second laser, which in a linear device must not depend on the wavelength or power of the first. Figure 4 summarizes this test for eight sessions (7 712 two-line waveforms).

The two parts of the photocurrent behave very differently. The DC photocurrent is linear and additive: the second laser's DC increment changes by only 0.3–1.4% when the first laser goes from 4 to 8 mW, and doubles when its own power doubles (Fig. 4b). The bias-modulated AC contribution is not additive. Figure 4(a) shows that the waveform contributed by the second laser changes visibly when the first laser goes from 4 to 8 mW; across sessions its amplitude grows by up to 31%, increasingly towards longer wavelengths (Fig. 4b). This coupling amounts to 21–35% of the contribution itself, 20–30 times the shot-noise floor and 5–20 times the change produced by shifting the line by 1 nm (Fig. 4c). The AC response is also nonlinear in power for a single line (Fig. 4d and Fig. 5). Writing the response at doubled power as $\Delta I(2P)=2^{\gamma}\Delta I(P)$, the effective exponent of the AC waveform varies smoothly and reproducibly with wavelength—from 0.94 at 1530 nm to 1.25 near 1560 nm, down to 0.17 near 1579 nm and back to 0.84 at 1610 nm—whereas the DC photocurrent stays at $\gamma\approx1.02$. Drift, power-calibration errors and oscilloscope clipping were ruled out as causes.

These observations show that the bias-modulated response is a property of the interaction between the photogenerated carriers and the swept internal field, which is absent from the DC photocurrent of the same device. The wavelength dependence of $\gamma$ itself carries spectral information, and because one line changes the response to another, a multi-line spectrum produces its own fingerprint rather than a sum of single-line fingerprints. This is what allows the spectrometer to recognize how many lines are present (Sec. 4.4), and it is why we learn the inverse mapping from calibration spectra, as in other nonlinear single-detector spectrometers [10,17], instead of inverting a responsivity matrix.

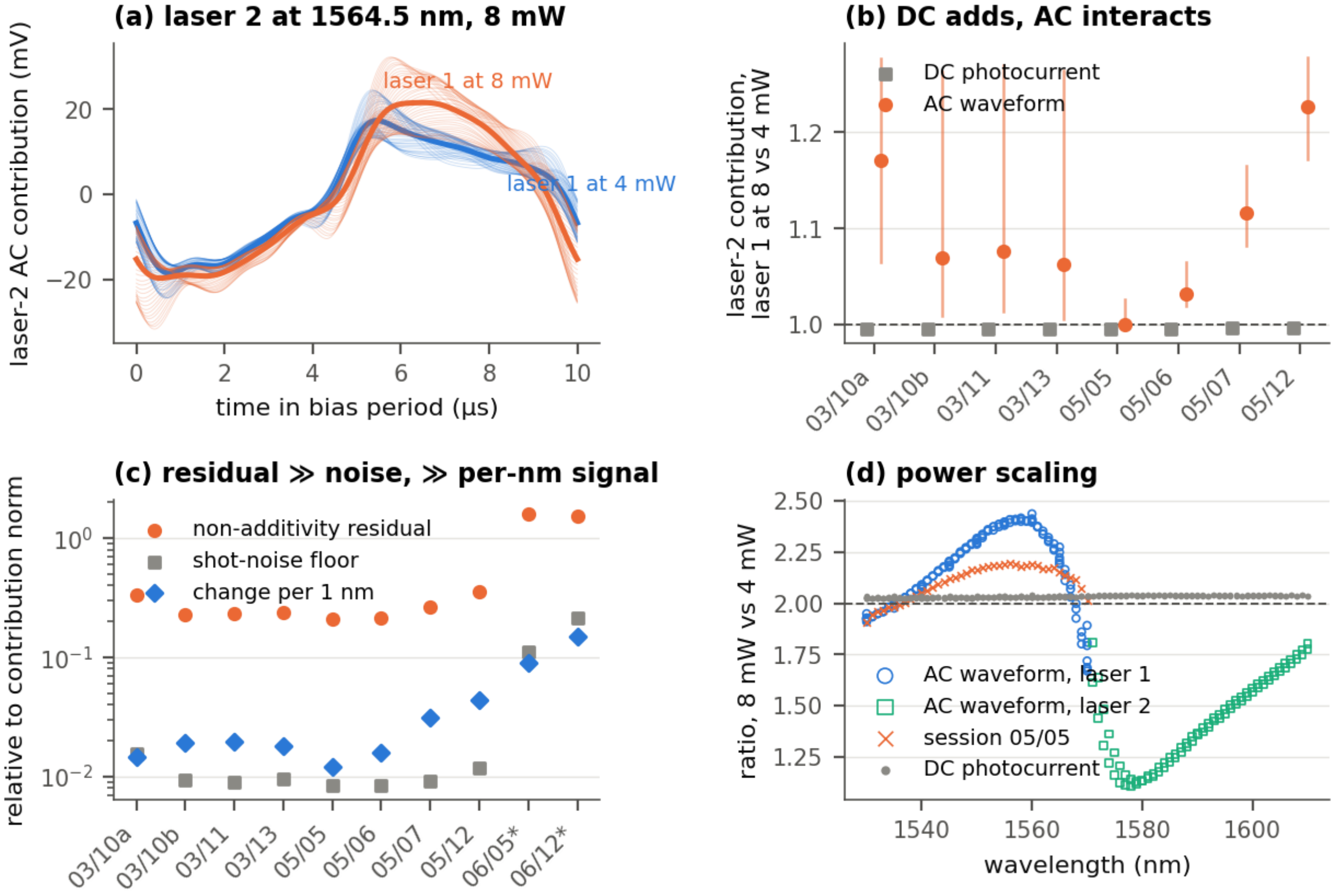

**Fig. 4.** Superposition test. (a) AC contribution of laser 2 (harmonics 1–8), measured as the difference between a two-line waveform and the single-line waveform recorded seconds earlier, with laser 1 at 4 (blue) or 8 mW (orange); thin lines: individual laser-1 wavelengths, thick: mean. (b) Ratio of laser 2's contribution with laser 1 at 8 vs 4 mW, per session: the DC part stays at 1 (additive), the AC part does not (bars: range over laser-2 wavelengths). (c) Coupling between lines compared with the shot-noise floor and with the change produced by a 1 nm step, relative to the contribution's magnitude; * June sessions, in which the single-line reference was recorded up to 160 measurements earlier. (d) Best-fit ratio of 8- to 4 mW waveforms versus wavelength for the AC waveform (open symbols; session 05_05 crosses) and the DC level (grey).

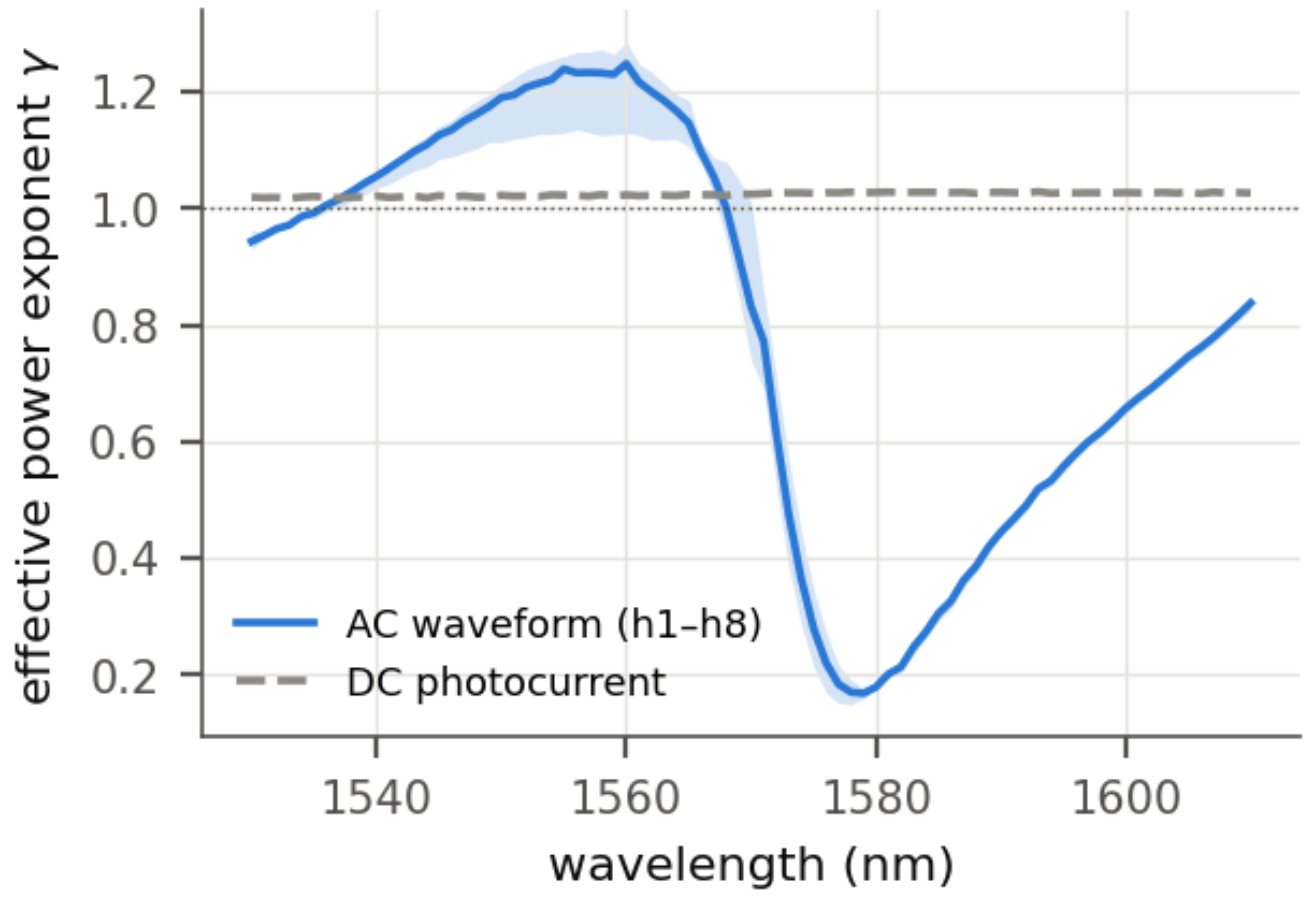

**Fig. 5.** Effective power exponent $\gamma=\log_2[\Delta I(8\text{ mW})/\Delta I(4\text{ mW})]$ of the bias-modulated photocurrent (harmonics 1–8; line: mean over sessions, band: range) and of the DC photocurrent, for single lines. A linear detector has $\gamma=1$.

### *4.3 Reconstruction*

Each dark-subtracted waveform is compressed onto 10–20 principal components, computed from calibration waveforms only, and a fully connected neural network maps them, together with the DC photocurrent, to the line wavelengths; a second network classifies the power of each line. For spectra with an unknown number of lines, a network with three output slots predicts for each slot whether a line is present, its wavelength and its power. The final model averages the wavelengths predicted from five complementary representations of the same waveform (principal components of the full and of the harmonic-filtered waveform, harmonic amplitude ratios and relative phases, DC-normalized harmonics, and the DC-normalized waveform). All model settings were chosen without the spectra on which accuracy is reported.

### *4.4 Multi-line spectral reconstruction*

The spectrometer is calibrated on part of the spectra measured in a set of sessions and tested on wavelengths or line combinations that were not used for calibration. For single lines, calibration uses only the 5 nm grid and the test covers the 64 wavelengths in between; for pairs and triples, no test combination appears in calibration.

**Wavelength accuracy.** Table 2 and Fig. 6 summarize the results. Single lines are located to 0.40 nm across the full 80 nm band, two simultaneous lines to 0.32 nm and three lines to 0.27 nm, with the power level of every single line and of both lines of 97% of the pairs identified correctly. Pairs separated by more than 20 nm are located to 0.26 nm, and the error grows smoothly as the lines approach (Fig. 6c). Averaging eight acquisitions of the same line improves the single-line accuracy to 0.23 nm.

**Table 2.** Wavelength accuracy (mean absolute error, nm) on wavelengths or line combinations not used for calibration. Single representation: principal components of the waveform; combined: five representations averaged (paired comparison, identical calibration sets).

| | Single representation | Combined | Change |
|---|---|---|---|
| 1 line | 0.42 | **0.40** | -5% |
| 2 lines | 0.36 | **0.32** | -11% |
| 3 lines | 0.34 | **0.27** | -22% |

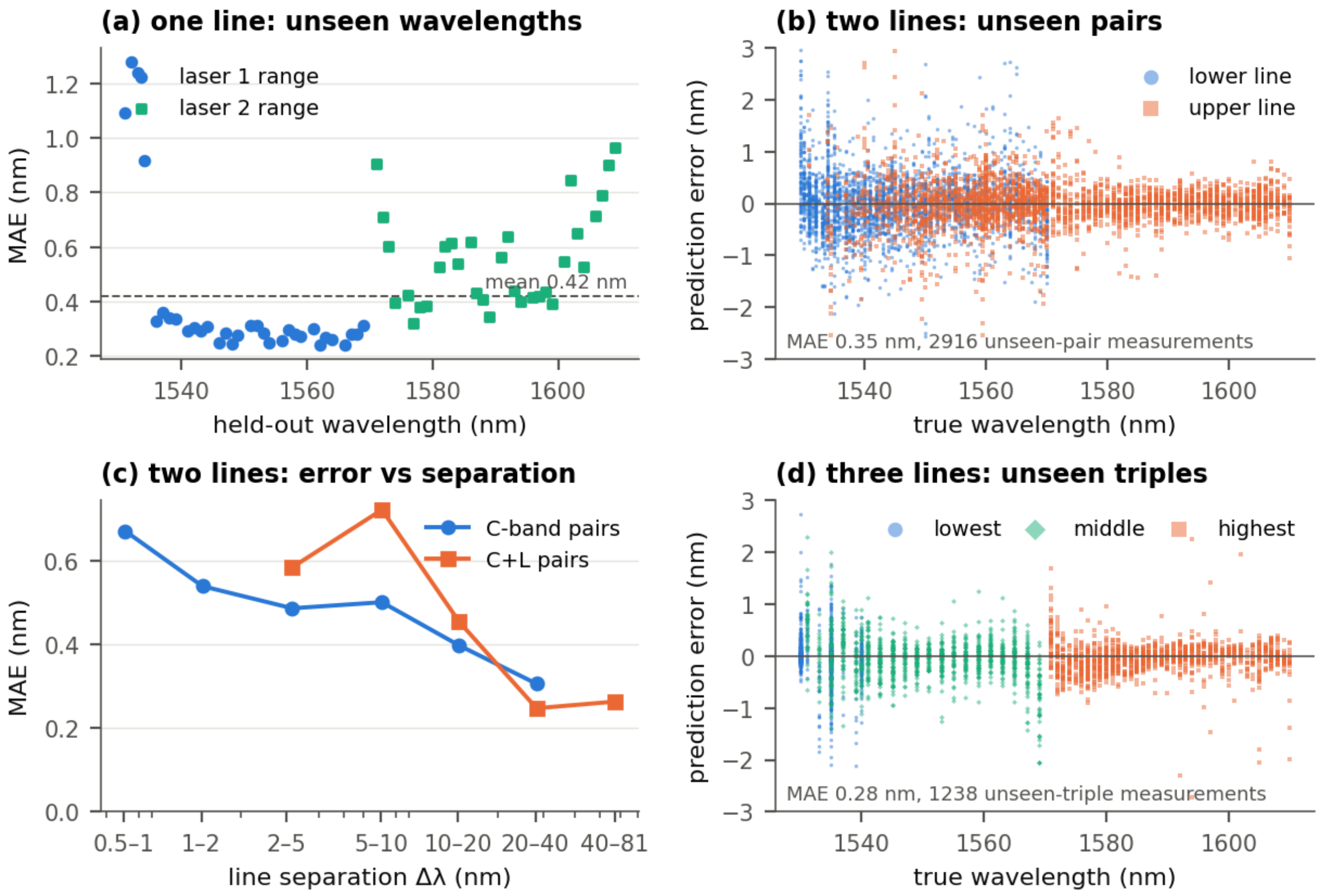

**Fig. 6.** Reconstruction with the number of lines known (single representation). (a) Single line: error per wavelength not used for calibration (calibration on the 5-nm grid). (b) Two lines: error of the lower and upper line for pairs absent from calibration. (c) Two-line error versus line separation for C-band and C+L pairs. (d) Three lines: error of each line for triples absent from calibration.

**Counting the lines.** A spectrometer should also tell how many lines are present. Without being told, the three-slot model identifies the number of lines in 98.0% of single-line spectra and 99.9% of pairs and triples (Table 3; examples in Fig. 7), locates single lines to 0.30 nm, pairs to 0.53 nm and triples to 0.46 nm, and identifies the power levels of single lines without error and of pairs in 97.5% of cases. Two lines only 0.5–1 nm apart are recognized as two lines in 98% of cases, giving a two-line resolution of about 1 nm. Consistent with the line coupling of Sec. 4.2, a linear inversion with a dictionary of measured single-line waveforms recovers single lines but not multi-line spectra (Table 3), confirming that the coupled response must be learned from calibration spectra of the same kind as those to be measured.

**Table 3.** Reconstruction without prior knowledge of the number of lines, on configurations not used for calibration: line-number accuracy, wavelength error for spectra with the correct number of lines, line $F_1$ score within one 1-nm bin, and power-level accuracy (one / two lines).

| Method | Number accuracy (1 / 2 / 3 lines) | Error, nm (1 / 2 / 3) | $F_1$ ±1 bin (1 / 2 / 3) | Power (1 / 2) |
|---|---|---|---|---|
| Three-slot model | 98.0 / 99.9 / 99.9% | 0.30 / 0.53 / 0.46 | 0.99 / 0.97 / 0.96 | 100 / 97.5% |
| Linear inversion (NNLS)[a] | – | – | – / 0.14 / 0.13 | – |

[a] Non-negative least squares with a dictionary of measured single-line responses covering every bin, number of lines given.

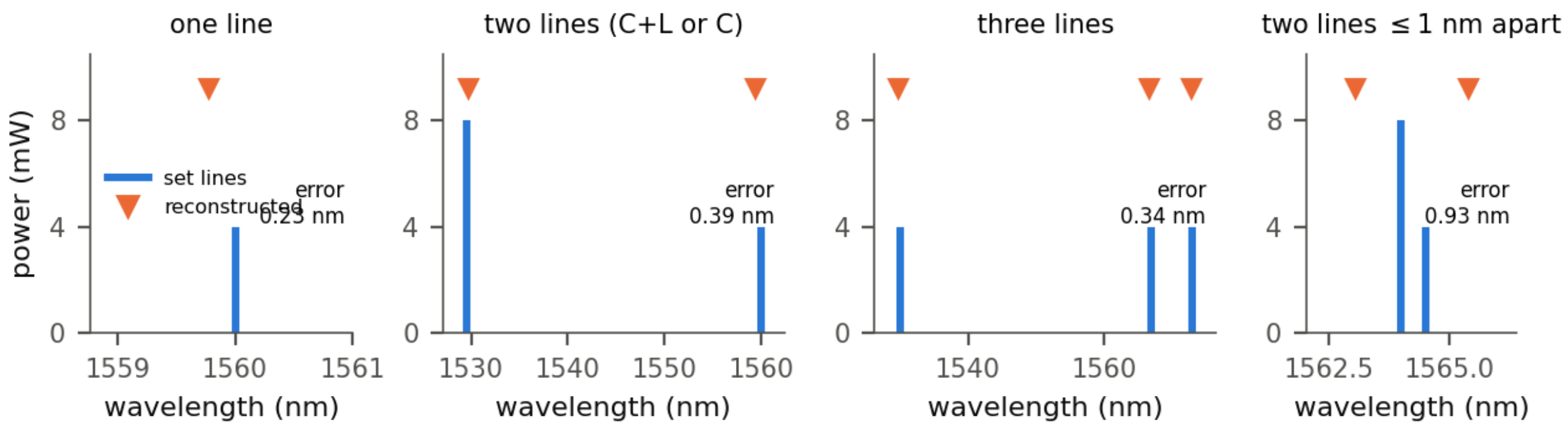

**Fig. 7.** Reconstructed spectra for configurations not used for calibration, without prior knowledge of the number of lines: set lines (blue, height = set power) and reconstructed wavelengths (orange triangles). Each panel is the test spectrum whose error equals the median of its family (one, two and three lines, and pairs at most 1 nm apart).

### *4.5 Tracking line shifts at the picometre scale*

Beyond locating lines over 80 nm, the waveform resolves much finer changes. Fine scans across 1550–1551 nm in 0.01 and 0.001 nm steps (Fig. 8) show that the waveform change per picometre relative to the noise supports a single-shot resolution of 28–58 pm. With a local calibration on a 0.01 nm grid, a single line is tracked to 48–60 pm on the intermediate steps, a second line moving 0.5–1 nm from a fixed one is tracked to 67 pm, and a calibration transferred to another day with three reference wavelengths still tracks the line to 57–63 pm. The direction of the waveform change agrees between sessions and with the coarse scans, confirming that the signal is wavelength. The same photodiode thus operates in two regimes: identification of lines across the C+L band, and picometre-scale tracking of line shifts.

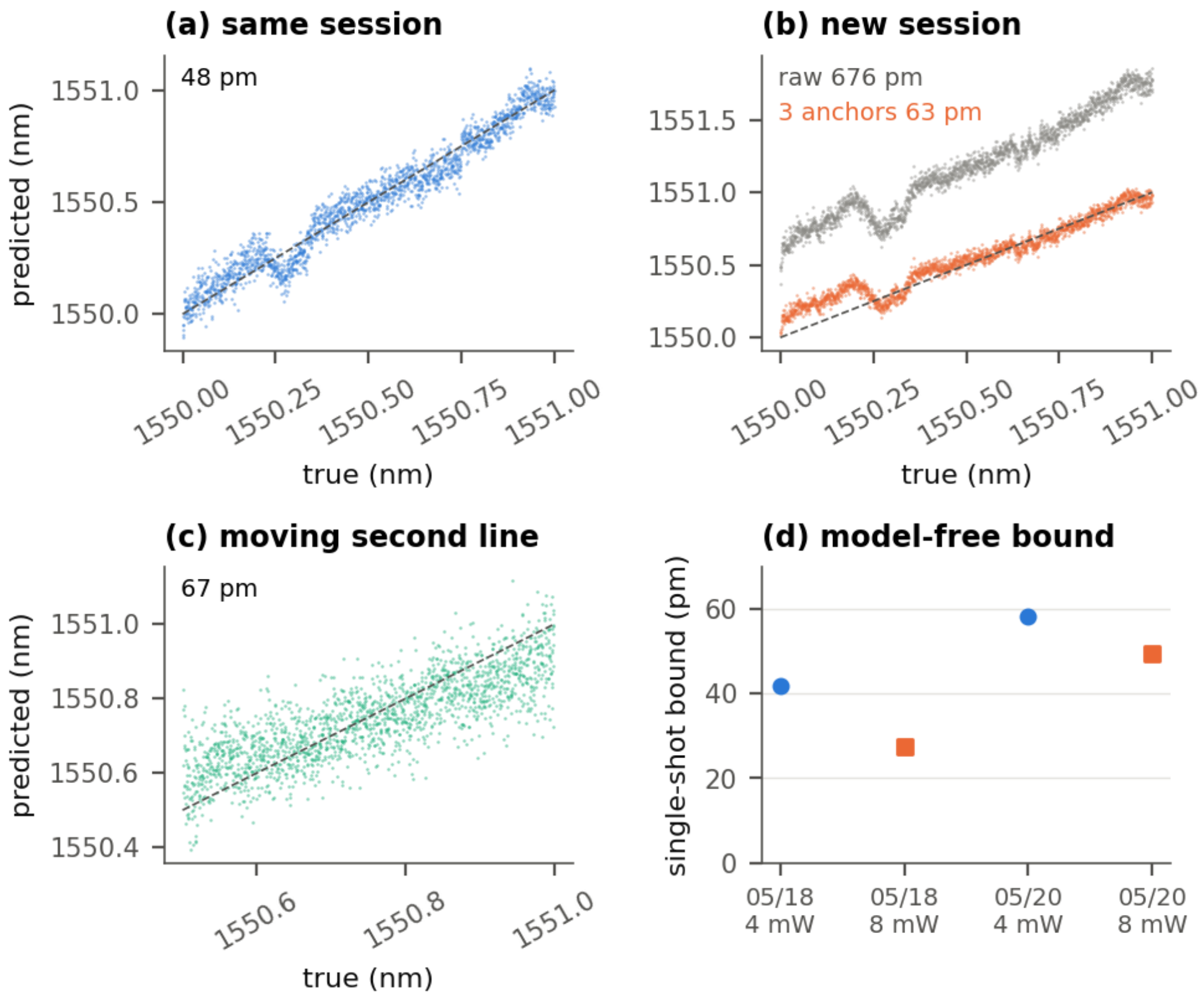

**Fig. 8.** Picometre-scale tracking with a linear estimator on harmonics 1–8. (a) Same session: calibrated on a 0.01 nm grid, tested on the intermediate 0.001 nm steps. (b) Model from session 05/18 applied to session 05/20, before (grey) and after (orange) three-point recalibration. (c) Second line moving 0.5–1.0 nm from a fixed line at 1550 nm, same session. (d) Model-free single-shot resolution bound (noise divided by the per-nm waveform change).

## 5. Discussion

**Performance without fabrication.** Table 4 compares the device with single-detector computational spectrometers. Every other entry depends on a detector or encoder fabricated for the purpose; ours is a commercial photodiode used as purchased. Despite this, its peak-wavelength accuracy of 0.27–0.40 nm lies in the range of the best engineered single-pixel devices (0.30 nm for a graded AlGaAs junction [14], 0.18 nm over 12 nm for a tandem organic detector [16]), and it holds over 80 nm for up to three simultaneous lines, where the engineered values were obtained with single lines. Its two-line resolution of about 1 nm matches the best reported values [7,16,17]. The device works at room temperature, in the telecom C+L band, and acquires the full photoresponse in one 10 µs bias period, whereas static-bias devices step through tens to about a hundred settings.

**Table 4.** Single-detector computational spectrometers and, below the line, an integrated photonic encoder. Values from the original papers; as compiled in Ref. [10]. –: not reported.

| Device | Fabrication of the spectral element | Range | Peak-wavelength accuracy | Resolution[a] | Encoding; reconstruction |
|---|---|---|---|---|---|
| Black phosphorus, Stark effect; 80 K [9] | custom 2D device | 2–9 µm | – | 90–420 nm | displacement field, 41–81 points; linear |
| $ReS_2$/Au/$WSe_2$ heterojunction [11] | custom 2D stack | 1150–1470 nm | – | ≈20 nm | gate, 25 points; linear |
| BP/$MoS_2$ tunnel diode [10] | custom 2D stack | 500–1600 nm | – | 2 nm | bias; linear |
| GeSe/InSe heterojunction [10,26] | custom 2D stack | 400–1100 nm | – | 0.35 nm | 101 bias points; neural network |
| $WSe_2$ PIN homojunction [10] | custom 2D device | 630–640 nm | 0.18 nm | 2 nm | bias and intensity; neural network |
| AlGaAs/GaAs p-graded-n junction [14] | custom graded epitaxy | 480–820 nm | 0.30 nm | 10 nm | depletion depth; neural field |
| Organic detector with optical spacer [15] | custom organic stack | 400–760 nm | – | ≈5 nm | generation depth, pulsed bias; linear |
| $MoS_2$ semi-floating homojunction [17] | custom 2D device | 450–800 nm | – | 1.2 nm | gate, amplitude and relaxation; neural network |
| Tandem organic photodetector [16] | custom organic tandem | 400–1000 nm | 0.18 nm (604–616 nm) | 1 nm | opposing subcells; linear |
| **This work**: InGaAs PIN, bias-modulated OED | **none (commercial, unmodified)** | 1530–1610 nm | 0.27–0.40 nm (1–3 lines); 50–60 pm tracking | ≈1 nm | generation depth, 100 kHz sinusoidal bias, one period; neural network |
| SiN interferometer cascade + InGaAs photodiode [7] | photonic integrated circuit | 1200–1700 nm | – | 1.4 nm | thermo-optic sweep, 200 points; Fourier |

[a]Minimum separation of two resolvable lines or width of a reconstructed narrow line. Ours: separation at which two laser lines are detected as two (Sec. 4.4).

**Operation on later days.** An instrument must work on days other than the day of its calibration. Stability tests of computational spectrometers have so far concerned repeated measurements, device uniformity, noise and temperature [7,8,15,17]; we go a step further and evaluate the spectrometer on entire measurement days excluded from calibration (Table 5, Fig. 9). On a new day the accuracy is set by slow day-to-day drift rather than by noise, and the drift is low-dimensional: a few waveform directions hold most of it (Fig. 3c), and it correlates with the session's dark reference, pointing to the bias and readout chain and to temperature. Because it is low-dimensional, it can be learned. Adding measured inter-day drift to the calibration set and averaging the five signal representations reduces the single-line error on a new day from 1.79 to 0.80 nm, with an improvement on every one of the seven test days, and to 0.61 nm when three reference lines are measured on the new day. Applied once, without further adjustment, to two fine-scan sessions that were never used during development, the same model reduced the error by 42% with three reference lines. For two-line spectra the new-day error is 0.8 nm; for three lines, recorded on fewer days, it remains above 1 nm, and calibration over more days, together with temperature stabilization of the photodiode and monitoring of the bias waveform, is the direct route to improving it.

**Table 5.** Accuracy on measurement days excluded from calibration (mean absolute error, nm; mean over test days; range in brackets). Uncalibrated: waveform only; recalibrated: plus three reference spectra on the test day.

| Task | Single representation | Combined + drift-aware calibration |
|---|---|---|
| 1 line, 7 days, uncalibrated | 1.79 [0.70–5.08] | **0.80** [0.34–1.93] |
| recalibrated | 1.18 | **0.61** [0.32–1.00] |
| 2 lines, 5 days, uncalibrated | 0.94 [0.83–1.06] | **0.82** [0.63–1.01] |
| 3 lines, 5 days, uncalibrated | 2.17 [1.09–3.36] | 1.92 [0.96–2.53] |
| recalibrated | 1.47 | 1.58 |

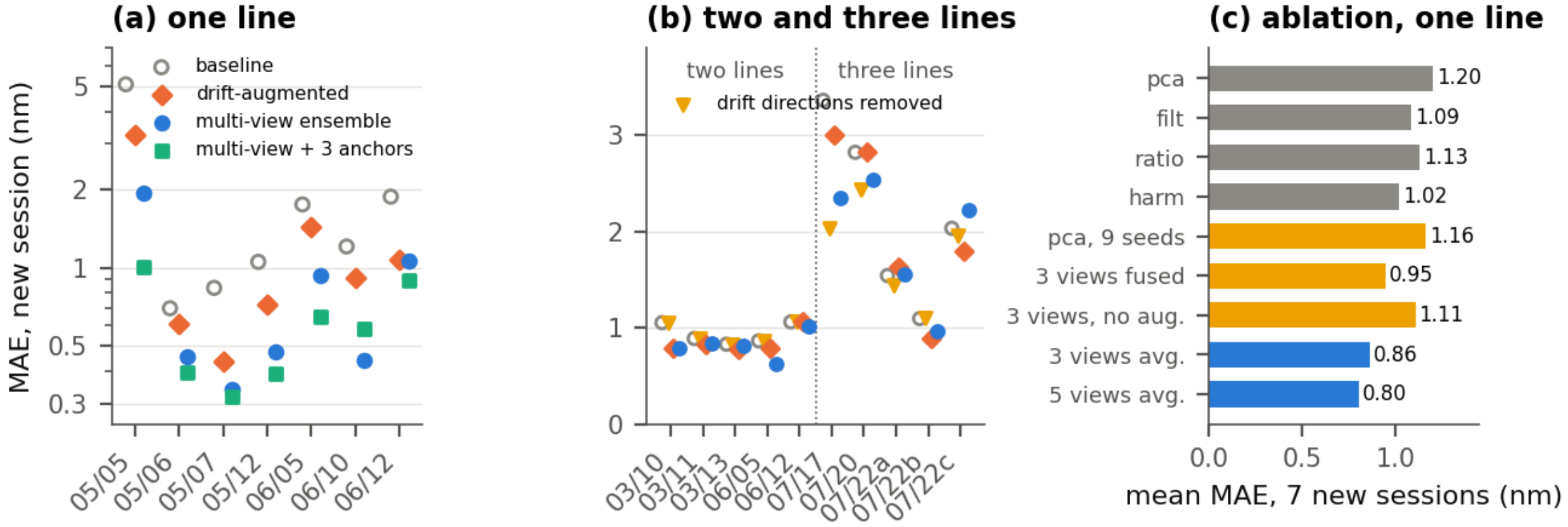

**Fig. 9.** Operation on days excluded from calibration. (a) Single line per test day: single representation (open grey), drift-aware single representation (orange), combined representations (blue) and combined with three-point recalibration (green). (b) Two-line (left) and three-line (right) test days; yellow triangles: drift directions removed. (c) Contribution of the ingredients for one line (mean over seven test days): single representations (grey), controls (yellow) and averaged representations with drift-aware calibration (blue).

**Physical origin.** Three observations constrain the mechanism: the spectral information lies in the first eight bias harmonics and near the bias crossings where $dV/dt$ is largest; the AC response is nonlinear in power with an exponent that varies across the band, while the DC photocurrent is linear; and the response to one line depends on the photocurrent generated by another. Together they point to an interaction between the photogenerated carriers and the bias-modulated internal field, for example space-charge screening at high photocurrent density (cf. bias-modulated OED in germanium photodiodes [25]). A drift–diffusion model of the PIN structure under the measured bias and power is the natural next step to make this connection quantitative.

**A generic, fabrication-free route to spectroscopy.** The experiments were performed on one InGaAs photodiode, but the mechanism is not specific to this device or material. It requires only (i) a wavelength-dependent absorption depth, (ii) a collection of carriers that depends on the applied bias, and (iii) a readout of the photocurrent waveform over a bias period. The first two are intrinsic to every semiconductor junction photodiode, and the third is standard electronics. OED has already been demonstrated in germanium photodiodes [19,20], silicon photodiodes and photovoltaic cells [21,23], and bias-modulated OED in germanium [25]. In principle, therefore, any off-the-shelf photodiode—silicon in the visible and near infrared, germanium and InGaAs in the telecom bands, extended InGaAs in the short-wave infrared—can be turned into a spectrometer for wavelengths just below its cutoff, where OED is strongest, by changing only its drive and its signal processing. Quantitative performance will depend on the absorber thickness, the doping profile and the bias range, and these become the design parameters, together with the bias frequency, amplitude and waveform shape, which can be chosen in electronics after the device is installed.

This opens a wide range of low-cost applications. Photodiodes are already present in every optical receiver, tap power monitor and transceiver; driving them with a bias waveform could add spectral functions without adding optical hardware: monitoring the wavelengths, number and powers of coarse wavelength-division-multiplexed channels, verifying and locking tunable lasers, or tracking laser drift and fiber-sensor signals with picometre-scale sensitivity, where OED has already enabled wavelength monitoring and fiber Bragg grating interrogation [20,22]. Silicon photodiodes, the cheapest optical sensors available, could likewise provide spectral sensing in the visible and near infrared for consumer, agricultural and industrial devices.

Natural next steps are the extension to lower optical powers and to broadband and continuous spectra, and the demonstration of the same procedure on photodiodes of other materials.

## 6. Conclusion

We have demonstrated that a standard commercial photodiode, used without any modification, becomes a spectrometer when it is driven by a sinusoidal bias. Bias-modulated optoelectronic chromatic dispersion encodes the spectrum in the shape of the photocurrent waveform over a single 10 μs bias period, and a neural network decodes it: one to three laser lines across the 80 nm of the telecom C+L band are located to 0.3–0.4 nm, their number and power levels are retrieved, and line shifts are tracked at the picometre scale. In contrast to existing single-detector computational spectrometers, the method requires no dedicated materials, no fabricated spectral structures, no optical filter and no modulation of the light, and it continues to operate on measurement days excluded from its calibration. Its generic, low-cost nature makes it readily adaptable to photodiodes of any material, turning detectors that today measure only optical power into fabrication-free spectral sensors and opening new directions for compact spectroscopy in optical communications, sensing and consumer devices.

## Funding

Israel Science Foundation (1215/26).

## Disclosures

The authors declare no conflicts of interest.

## Data availability

Data underlying the results presented in this paper are available from the authors upon reasonable request.